\documentclass{PoS}
\usepackage{natbib}
\usepackage{epsfig}
\usepackage{amsmath}
\usepackage{amssymb}
\usepackage{latexsym}
\usepackage{subfigure}
\usepackage{natbib}
\usepackage{txfonts}

\newcommand\arcmin{\mbox{$^\prime$}}
\PoS{PoS(BDMH2004)065}

\title{The Outer Cluster System of NGC\,1399: Preliminary Results}

\ShortTitle{The Outer Cluster System of NGC\,1399: Preliminary Results}

\author{\speaker{Ylva Schuberth}\\
  Sternwarte der Universit\"at Bonn,  Auf dem H\"ugel 71, D-53121 Bonn, Germany\\
        E-mail: \email{ylva@astro.uni-bonn.de}}
\author{Tom Richtler\\
  Universidad de Concepci\'on, Departamento de F\'isica, Casilla  160-C, Concepci\'on, Chile\\
        E-mail: \email{tom@coma.cfm.udec.cl}}
\author{Boris Dirsch\\
  Universidad de Concepci\'on, Departamento de F\'isica, Casilla  160-C, Concepci\'on, Chile\\
         E-mail: \email{bdirsch@cepheid.cfm.udec.cl}}
\author{Michael Hilker\\
  Sternwarte der Universit\"at Bonn,  Auf dem H\"ugel 71, D-53121 Bonn, Germany\\
        E-mail: \email{mhilker@astro.uni-bonn.de}}
\author{S{\o}ren Larsen\\
  European Southern Observatory, Karl-Schwarzschild-Str.~2, D-85748 Garching, Germany \\
        E-mail: \email{slarsen@eso.org}}

\abstract{We present preliminary results of our dynamical study of the
          outer globular cluster system of NGC\,1399, the central
          galaxy in the Fornax cluster. About 160 new radial
          velocities for globular clusters at projected galactocentric
          distances between 8\arcmin\, and 18\arcmin\, indicate that
          the constant velocity dispersion of $\sim276\,\textrm{km\,s}^{-1}$
          (for all clusters) already found for the inner region can be
          traced out to 80\,kpc. We find that the kinematical
          properties of the blue (metal-poor) and the red (metal-rich)
          globular cluster subpopulations appear to be 
          different: While the velocity distribution of the red
          clusters is symmetric with respect to the systemic velocity
          of NGC\,1399, the blue clusters show a somewhat asymmetric
          distribution, with more velocities above the systemic
          velocity.}

\FullConference{Baryons in Dark Matter Halos\\
		 5-9 October 2004\\
		 Novigrad, Croatia}

\begin{document}

\section{Introduction}
\begin{figure}
\epsfig{file=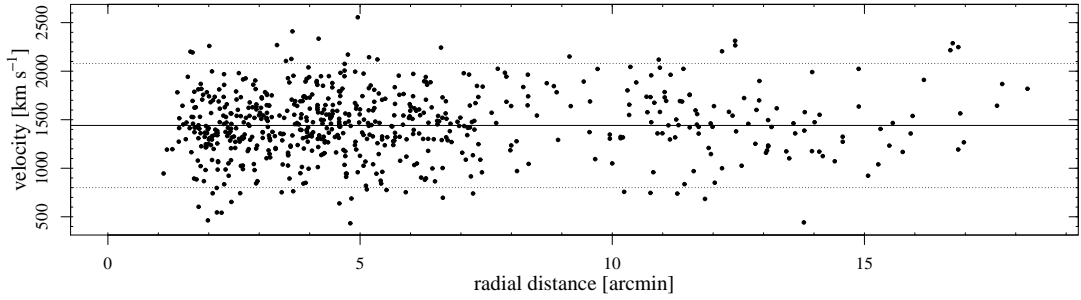, width=.99\textwidth,angle=0}
\caption{Radial velocity versus projected galactocentric radius for the
combined sample. The solid line at $1441\,\textrm{km\,s}^{-1}$ indicates the
systemic velocity of NGC\,1399. The dotted lines at $800$ and $2080\,
\textrm{km\,s}^{-1}$ show the velocity selection we adopted from Richtler et
al.~(2004).}
\label{scatter}
\end{figure}

NGC\,1399, the central giant elliptical in the Fornax Cluster has long
since been known to host a very populous and extended globular cluster
system (GCS). In the Southern hemisphere it is the prime target when
using globular clusters as probes for the gravitational potential of
the host galaxy.  Recent wide-field photometry in the Washington
system (Dirsch et al.~2003) provided the database to select GC
candidates for the kinematical and dynamical study of the NGC\,1399
GCS presented by Richtler et al.~(2004) and Dirsch et al.~(2004).
They used the VLT with FORS2/MXU to obtain a sample of 470 GC
velocities in a radial range of $2\arcmin < R < 9\arcmin$ (11 to
50\,kpc in a distance of 19 Mpc), the largest sample of GC velocities
measured until then. Briefly, their main findings are: For the
entire sample, they find a radially constant velocity dispersion of
$\sigma=325\pm 11\,\textrm{km\,s}^{-1}$.  Omitting the extreme velocities,
i.e.~taking into account only the clusters within the velocity
interval $800 < \varv < 2080\, \textrm{km\,s}^{-1}$ (see
Fig.~\ref{scatter}), they find $\sigma_{\rm{all}}=276\pm
11\,\textrm{km\,s}^{-1}$.  Inspecting blue (metal-poor) and red (metal-rich)
clusters separately, the corresponding dispersions read:
\mbox{$\sigma_{\rm{blue}}= 291\pm 14\,\textrm{km\,s}^{-1}$} and
\mbox{$\sigma_{\rm{red}}= 255\pm 13\,\textrm{km\,s}^{-1}$}.  Using a
dynamical model on the basis of the spherical Jeans equation, they
derive -- under the assumption of isotropy -- a radially constant
circular velocity of \mbox{$\varv_{\rm{circ}}=415\pm
30\,\textrm{km\,s}^{-1}$} out to a radius of 50\,kpc.

It is of great interest to extend this study to larger galactocentric
distances, in order to address the several key questions that can put
constraints on galaxy formation scenarios: Out to what radius can the
constant circular velocity be followed?  Does the outer GCS rotate?
Can one detect substructure in the dark matter distribution of the
Fornax cluster (e.g.~Ikebe et al.~1996)?  Here we present the first
preliminary results from a study measuring cluster velocities at
projected galactocentric radii of up to 100\,kpc.
\section{The Data Set}
The data have been obtained with FORS2 and the Mask Exchange Unit
(MXU) at the Very Large Telescope of the European Southern Observatory
at Cerro Paranal, Chile. The observing period was December 2/3, 2002
(ESO program ID 70.B-0174). Twelve masks in 7 different fields have
been observed.  The spectral resolution provided by the grism 600B is
about 3\AA.  The data structure and the reduction procedure are
identical to those in our previous study of the inner cluster system
(Richtler et al.~2004, Dirsch et al.~2004). The data are not yet
completely reduced. So far, we have determined 160 new GC velocities
with typical uncertainties of $30\,\textrm{km\,s}^{-1}$. For 139 of these
objects, colours are available from the Washington photometry of
Dirsch et al.~(2003).
\begin{figure}
\epsfig{file=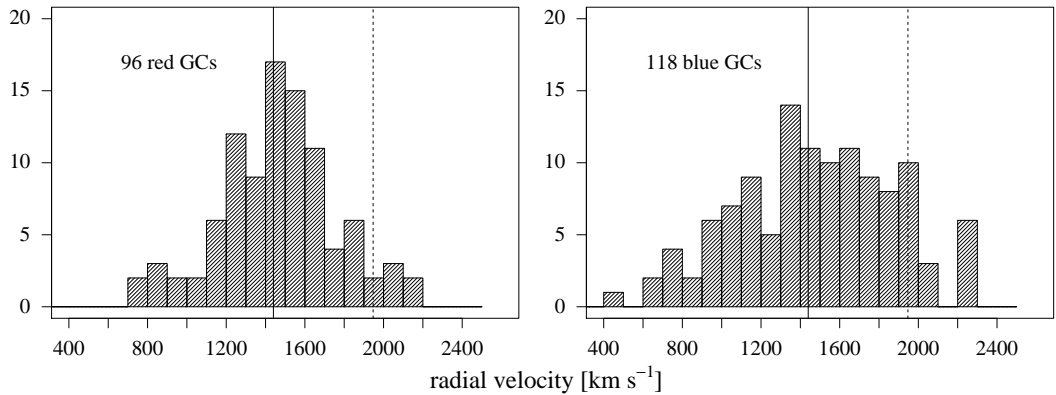, width=.99\textwidth,angle=0}
\caption{Histograms of the velocity distribution for the outer $(R >
 5.5\arcmin)$ red and blue clusters. The solid line marks the systemic
 velocity of NGC\,1399. The dotted line at $1950\,\textrm{km\,s}^{-1}$
 indicates the systemic velocity of NGC\,1404.}
\label{distri}
\end{figure}
\begin{figure}
\centering
\epsfig{file=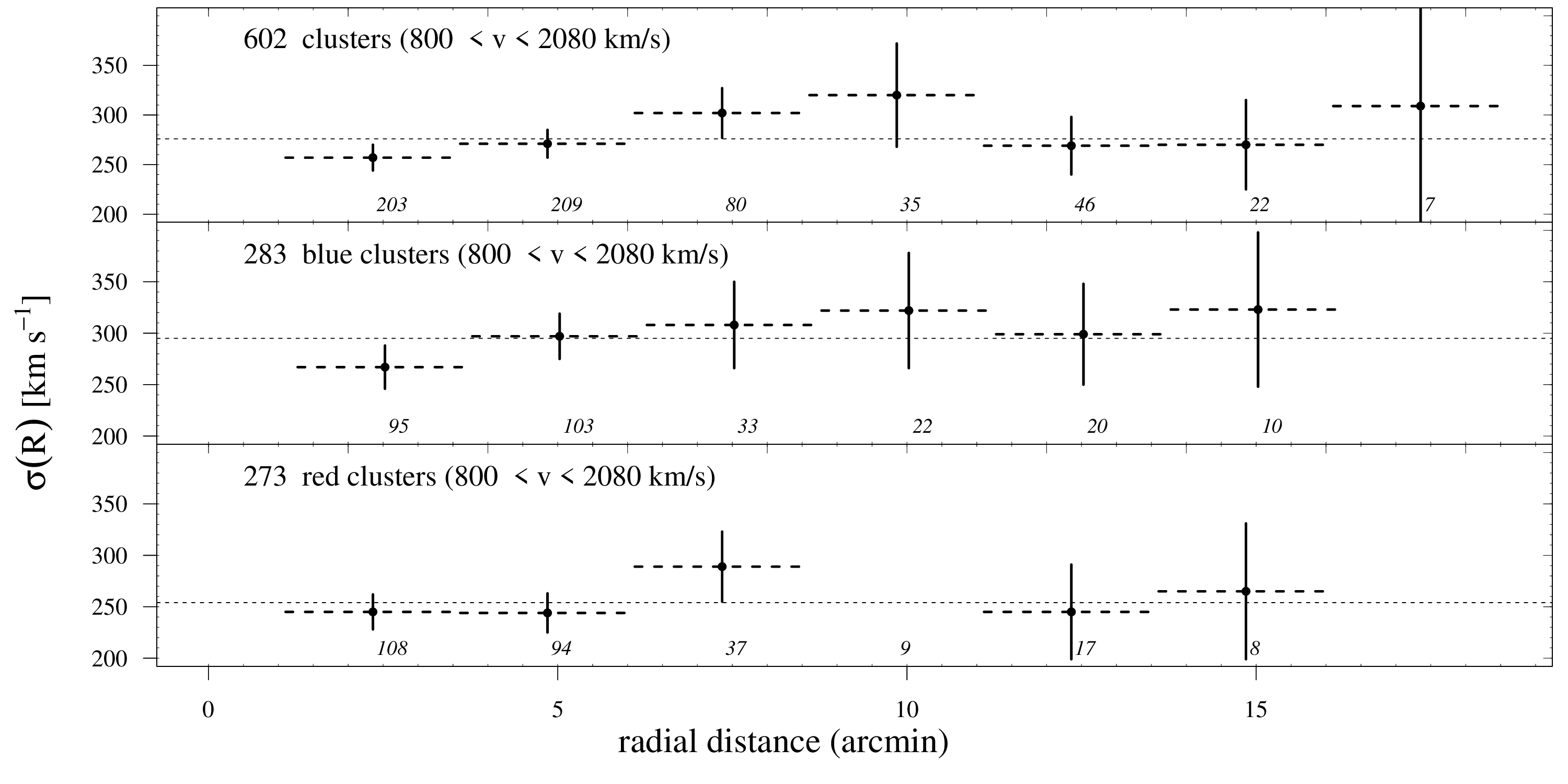, width=.99\textwidth,angle=0}
\caption{Velocity dispersion as function of projected radius. The
upper panel shows the values for all selected clusters. The results
for blue and red clusters are shown in the middle and bottom panel,
respectively. In all panels, the dotted line indicates the mean of the
corresponding dispersion measurements. The number of clusters entering
the dispersion measurements for a given bin of 2.5\arcmin\, width is
indicated as well.}
\label{disprad}
\end{figure}

\section{Results}

Figure~\ref{scatter} shows a plot of the new radial velocities
versus galactocentric distance together with the older data. Indicated
are the systemic velocity and the boundaries within which we determine
the velocity dispersion to be consistent with the previous
selection. The outermost data point has a projected radial
distance of 18\arcmin, corresponding to 100\,kpc. Given that more
velocities will soon follow, the velocity
dispersions reported here might still change somewhat in the future.

A peculiarity is apparent in the velocity distribution of the outer
clusters shown in Fig.~\ref{distri}. While the red clusters show a
reasonably symmetric distribution, the outer blue clusters exhibit an
asymmetry, with a preference for higher velocities. A peak at
$1800\,\textrm{km\,s}^{-1}$ was already found for the blue clusters at
smaller radii in the sample of Richtler et al.~(2004).  These findings
are interesting in the context of the simulations presented by Bekki
et al.~(2003) who considered an interaction between the nearby
elliptical NGC\,1404 and NGC\,1399, an idea first brought forward by
Kissler--Patig et al.~(1997).  In the scenario of Bekki et al., where
NGC\,1404 is on a bound orbit around the Fornax cluster centre
(i.e. around NGC\,1399), the low specific frequency (the total number
of clusters normalised to the host galaxy's luminosity) of NGC\,1404
is explained by the tidal stripping of NGC\,1404 clusters which
subsequently form an additional GC population around NGC\,1399. This
mostly affects the blue clusters in NGC\,1404 due to their shallower
number density profile.  If true, this would complicate the use of the
blue clusters as tracers for the gravitational potential of NGC\,1399.
However, more velocities are needed to arrive at safer conclusions.

The projected velocity dispersions derived for the total, the blue,
and the red sample are shown in Fig.~\ref{disprad}. The horizontal
dotted lines indicate the mean value of the respective dispersions and
are in excellent agreement with the previously derived values for the
inner region. Apparent dissimilarities, for instance the slight rise
of the velocity dispersion of the blue (and the total) sample may be
due to differences in the binning. The highest velocity dispersion
values (at $10\arcmin$) are probably not reliable since they trace a
radial interval where, as a comparison with Fig.~\ref{scatter} shows, the
distribution of velocities is highly asymmetric. \par At present, the
global picture is that of a constant velocity dispersion over the full
radial range.  The total sample has a velocity dispersion of
$276\,\textrm{km\,s}^{-1}$.  The dispersion for the red clusters is
$254\, \textrm{km\,s}^{-1}$ and for the blue clusters $295\,
\textrm{km\,s}^{-1}$.  It also seems that the difference between the
blue and the red clusters found for the inner region, remains in the
outer region. This is remarkable since the number density profiles of
both populations are indistinguishable for radii beyond $8\arcmin$\,
(Dirsch et al.~2003).  In case of isotropy, the blue clusters should
assume the same velocity dispersion as the red clusters beyond this
radius.  Summarising, the present data support the presence of an
isothermal dark halo which extends to at least 80\,kpc.

\end{document}